\documentclass[
aip,
pof,
amsmath,amssymb,
preprint,%
]{revtex4-2}

\usepackage{graphicx}
\usepackage{soul,color}
\usepackage{dcolumn}
\usepackage{bm}

\begin{document}

	
	\title{The role of atomization in the coupling between doped droplets dynamics and their flames}   

	\author{Sepehr Mosadegh and Sina Kheirkhah}
	\email[Author to whom correspondence should be addressed:~]{sina.kheirkhah@ubc.ca}	
	\affiliation{School of Engineering, The University of British Columbia, Kelowna, British Columbia, Canada, V1V 1V7}


\begin{abstract}
The droplet and flame chemiluminescence dynamics as well as their coupling during atomization events of graphene oxide doped diesel are investigated experimentally. The tested doping concentrations are 0, 0.001, 0.005, 0.01, and 0.02\% by weight. To minimize heat transfer between the droplet and its suspension mechanism, small diameter fibers are used for the droplet suspension. Separate shadowgraphy and $\mathrm{OH^*}$ chemiluminescence measurements are performed at 4000~Hz to study the droplet and flame dynamics, respectively. The results show that both the droplet diameter squared and the flame chemiluminescence feature intermittent oscillations. The droplet diameter squared oscillations RMS is positively related to the number and intensity of the atomization events and the graphene oxide doping concentration. The probability density function of the inverse of the time separation between two consecutive atomization events and the power spectrum density of the droplet diameter squared oscillations feature dominant large probabilities and powers at about 25~Hz prior to the occurrence of the first intense atomization event. After the occurrence of the first intense atomization event, this frequency decreases to about 5~Hz for both. Although the intense atomization triggers the large amplitude oscillations at 5~Hz, it was argued that the retracting motion of the igniter induces the oscillations at 25~Hz. Our findings suggest that the atomization events are the root cause of the smaller frequency coupling between liquid fuel droplets doped with graphene oxide and their flames. This has implications for spray combustion research.
\end{abstract}
	
\maketitle

\section{\label{sec:Introduction}Introduction}
	
As doping agents, nanomaterials have been used to improve several liquid fuels combustion characteristics, such as their burning rate and emissions, see for example~\cite{sabourin2009functionalized,tanvir2016droplet,sim2018effects,javed2013evaporation,ghamari2017combustion,mosadegh2021graphene}. In the context of droplet combustion, it is shown that doping of liquid fuels with nanomaterials leads to the occurrence of the atomization phenomenon~\cite{ghamari2017combustion,mosadegh2021graphene,basu2016combustion,gan2011combustion,miglani2015coupled,pandey2019high,yao2021atomization}. This changes the spatio-temporal size distribution of the droplets and their burning rate~\cite{mosadegh2021graphene}, which may influence the dynamics of coupling between the droplets and the corresponding flames. Our understanding related to the characteristics of the droplet-atomization-flame coupling is limited, and this requires further investigations.
	
The atomization phenomenon is comprised of the nucleation and bubble growth, droplet surface rupture, ligament formation and its break up, and ejection of baby droplets \cite{basu2016combustion}. Nucleation occurs at the liquid-liquid interface of multicomponent fuels with different boiling temperatures, at the liquid-solid interface of the droplet and its suspending mechanism, and/or at the nanomaterials surface~\cite{law1982recent,basu2016combustion}. For the latter, it is shown that the addition of carbonaceous nanomaterials to liquid fuels allows for absorption of the combustion heat at the materials surface, formation of nucleation sites, and as a result generation of bubbles at these sites~\cite{ghamari2017combustion,mosadegh2021graphene}. Coalescence of the bubbles, transfer of heat from the flame to the bubbles, and the surface regression of the main droplet lead to formation of relatively large bubbles~\cite{pandey2018boiling,ghamari2017combustion,javed2013evaporation}. Once these large bubbles reach the droplet surface, it is ruptured, and ligaments are formed~\cite{basu2016combustion,rao2018bubble}. Finally, the ligaments break up and baby droplets are ejected~\cite{pandey2019high}.
	
The atomization intensity is highly influenced by the concentration and the type of the doping agents. For moderate doping concentrations, the nanomaterials form a shell that is continuously ruptured during the atomization events~\cite{miglani2015coupled}. However, for excessively large concentrations of the nanomaterials, the shell is relatively strong, which traps the bubbles, and suppresses the atomization. Mosadegh \textit{et al.}~\cite{mosadegh2021graphene} showed that the amount of oxygen available in the graphene oxide molecular structure can also influence the atomization phenomenon. It was reported that partially oxidized graphene oxide features a relatively large infrared absorption which was discussed to increase the probability of the occurrence of the atomization~\cite{mosadegh2021graphene}.
	
The droplet burning rate is defined~\cite{law1982recent} as the negative of the derivative of the main droplet diameter squared ($D^2$) with respect to time ($t$). Mosadegh \textit{et al.}~\cite{mosadegh2021graphene} showed that, for graphene oxide doped ethanol droplets, variation of $D^2$ versus $t$ can be treated separately for time periods before and after the first atomization event. For the time period prior to the occurrence of the first atomization event, the burning rate corresponds to the mass loss due to the droplet evaporation; and, this rate increases with increasing the thermal conductivity of the doped fuel. However, the burning rate during the atomization period features contributions from both the evaporation and ejection of the baby droplets. It was shown~\cite{mosadegh2021graphene} that the burning rate during the atomization period is positively correlated with the infrared absorption of the doping agent. Mosadegh \textit{et al.}~\cite{mosadegh2021graphene} studied the characteristics of the baby droplets ejections and showed these occur during the second half of the droplet lifetime. It was concluded that the most expected baby droplet diameter and ejection velocity are about 50~$\mu$m and 0.5~m/s, respectively. They \cite{mosadegh2021graphene} estimated the mass of the ejected baby droplets during the atomization period, and showed that the main droplets which eject relatively large mass of the baby droplets feature relatively small evaporation mass loss. Although the influence of the nanomaterials doping on the droplet burning rate is of significant importance, it is unclear how atomization impacts the droplet and flame coupling.
	
Pandey \textit{et al.}~\cite{pandey2019high} analyzed the droplet shape deformation and the flame chemiluminescence oscillations. For ethanol mixed with water, they~\cite{pandey2019high} reported that the droplets shape deformation and their flame chemiluminescence feature oscillations at about 32 and 43~Hz, respectively. They showed that the addition of nanoceria particles to the droplets decreases both of the above frequencies to 13~Hz. Their results~\cite{pandey2019high} suggest that the droplet shape deformation and the flame chemiluminescence oscillations may be coupled through two mechanisms. First, the atomization events induce the droplet shape deformation. This changes the fuel vapour flux into the flame and as a result the flame chemiluminescence. Second, the atomized baby droplets are ignited as they reach the flame, and this induces alteration of the flame chemiluminescence. Although the results of Pandey \textit{et al.}~\cite{pandey2019high} are of significant importance as they elaborate the mechanisms of the droplet shape deformation and the flame chemiluminescence oscillations, the root cause of these oscillations is unclear, and this requires further investigations.
	
The first objective of the present study is to investigate if the coupling observed in Pandey \textit{et al.}~\cite{pandey2019high} for nanoceria doped ethanol droplets also exists for graphene oxide doped diesel droplets. Our second objective is to discover the root cause of such coupling and to investigate the effect of graphene oxide doping concentration on the above coupling.
	
\section{\label{sec:Methodology}Experimental methodology}

Details of the doped fuel preparation and tested conditions, experimental setup, diagnostics, and data reduction are presented below.
	
\subsection{\label{Sec:Nanofuel}Doped fuel preparation and the tested conditions}
Graphene oxide (GO) functionalized with octadecyl amine groups was used as the doping agent in the present study. The chemical composition of the tested GO was investigated using the Energy-Dispersive X-ray Spectroscopy. The results showed that the tested GO platelets are mainly comprised of about 80\% and 20\% by weight of carbon and oxygen, respectively. The GO nanomaterials were suspended in ultra-low sulphur diesel, and the suspension was sonicated for 60~min prior to each experiment. No surfactant was used during the suspension preparation. In total, five doping concentrations of 0, 0.001, 0.005, 0.01, and 0.02\% (by weight) were examined, which are denoted by D, D001, D005, D01, and D02, respectively, and are presented in Table~\mbox{\ref{tab:Conditions}}.
	
\begin{table} [h!] 
		\caption {The doped fuel tested conditions. \label{tab:Conditions}}
		\centerline{\begin{tabular}{c c c c c c}
				\hline
				Condition~~~~~~~~~~ & ~~D & ~~D001 & ~~D005 & ~~D01 & ~~D02  \\
				\hline
				$\mathrm{[GO]}$(\% by weight)~~~~~~~~~~ & ~~0 & ~~0.001 & ~~0.005 & ~~0.01 & ~~0.02 \\
				\hline	\end{tabular}}
\end{table}
	
\begin{figure}[h!]
\centering
\includegraphics[width=0.5\textwidth]{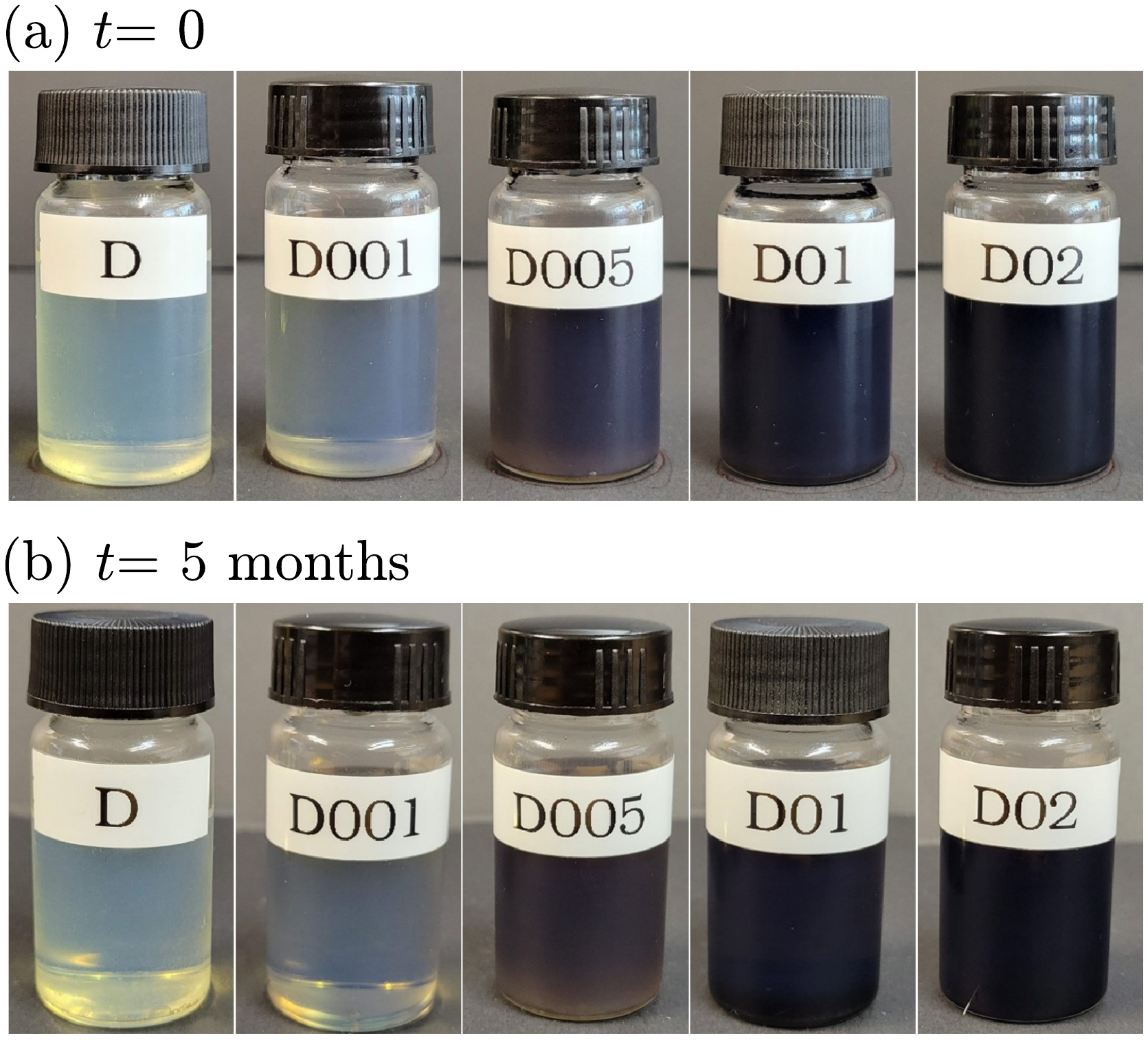}
\caption{Photographs of the doped fuels (a) immediately after preparation, and (b) 5 months after preparation.
\label{fig:Stability}}
\end{figure}

To assess the stability of the suspensions, their photographs were acquired. The photographs corresponding to neat diesel, D001, D005, D01, and D02 are presented in the first to fifth columns of Fig.~\ref{fig:Stability}, respectively. The images presented in Figs.~\ref{fig:Stability}(a)~and~(b) correspond to those collected immediately after preparation of the fuel as well as 5 months later. As can be seen, the doped fuels were stable and no sedimentation was observed during the above period.
	
\begin{figure}[h!]
	\centering
	\includegraphics[width=0.5\textwidth]{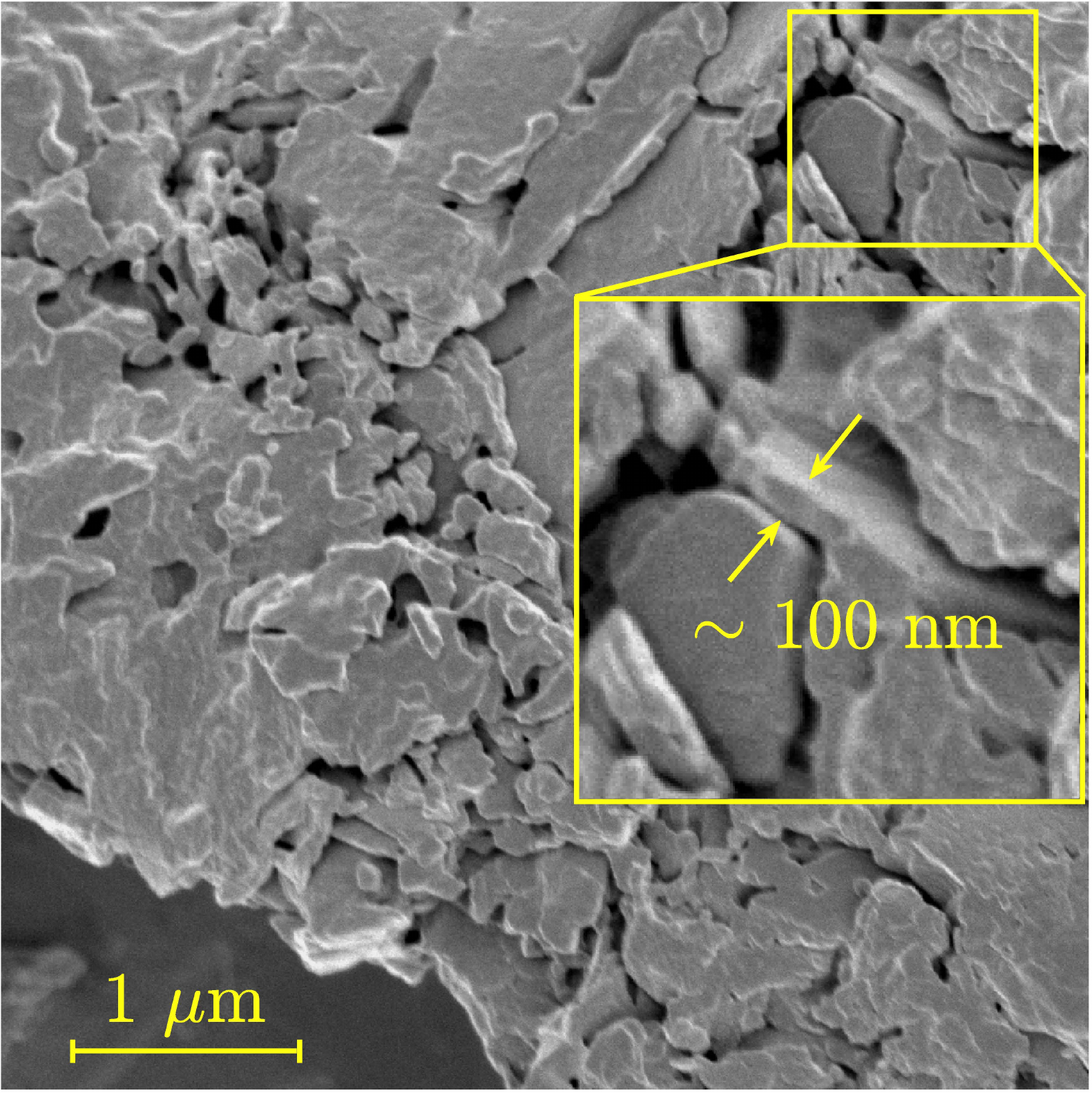}
	\vspace{1 pt}
	\caption{Scanning electron microscopy of the stacked GO. 
		\label{fig:SEM} }
\end{figure}

\begin{figure}[h!]
		\centering
		\includegraphics[width=0.5\textwidth]{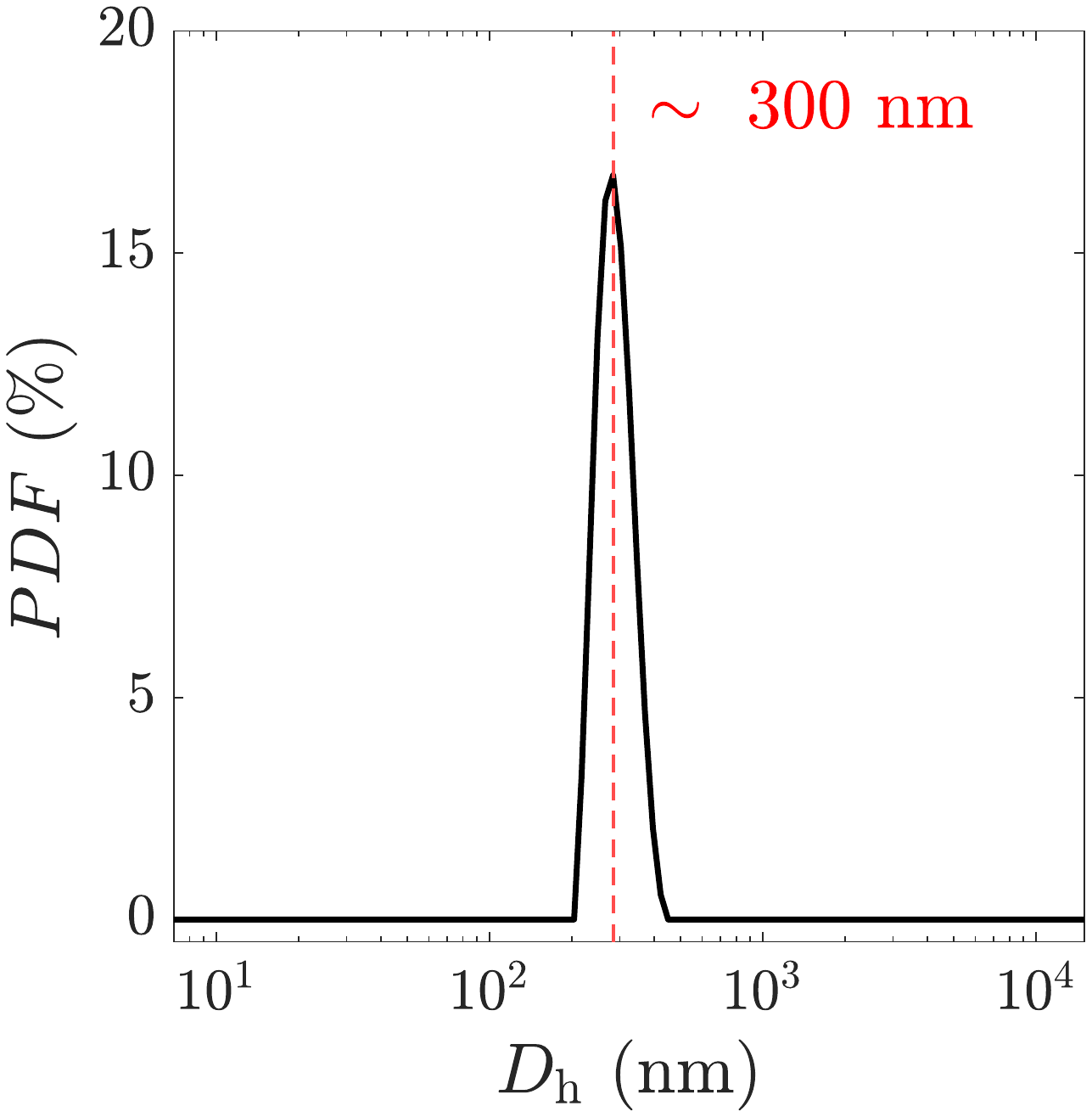}
		\vspace{1 pt}
		\caption{Probability Density Function of the GO hydrodynamic diameter for the test condition of D02. 
			\label{fig:DLS} }
\end{figure}

The Scanning Electron Microscopy (SEM) was performed to study the morphology of the GO. Figure~\mbox{\ref{fig:SEM}} shows an SEM image of the stacked GO aggregates after letting the diesel in the doped fuel to completely evaporate. Despite that the size of the GO platelets suspended in liquid fuel cannot be measured from such SEM (due to aggregation of the platelets), the GO sheet thickness can be obtained and was about 100~nm, see the inset of Fig.~\mbox{\ref{fig:SEM}}. To estimate the hydrodynamic diameter ($D_\mathrm{h}$) of the GO platelets suspended in diesel, the Dynamic Light Scattering (DLS) experiments were conducted using a NanoLab 3D. This device utilizes a 638~nm light source. The temperature of GO and diesel suspension and the scattering angle of the DLS measurements were 25$\mathrm{^o}$C and 150$\mathrm{^o}$, respectively. The DLS experiments were conducted for a run duration of 45~s, and each run was repeated 10 times. The hydrodynamic diameter of the GO platelets was obtained using the method of cumulants discussed in Hassan \textit{et al.}~\mbox{\cite{hassan2015making}}. The Probability Density Function ($PDF$) of the hydrodynamic diameter for a representative test condition of D02 was obtained and is presented in Fig.~\mbox{\ref{fig:DLS}}. The results showed that the GO platelets suspended in diesel feature a most probable hydrodynamic diameter of about 300~nm, see Fig.~\mbox{\ref{fig:DLS}}. Similar results were obtained from the rest of the tested conditions.
	
\subsection{Setup and diagnostics} 
	\label{Sec:ExpDigDat} 
\begin{figure*}[t!]
	\centering
	\includegraphics[width=1\textwidth]{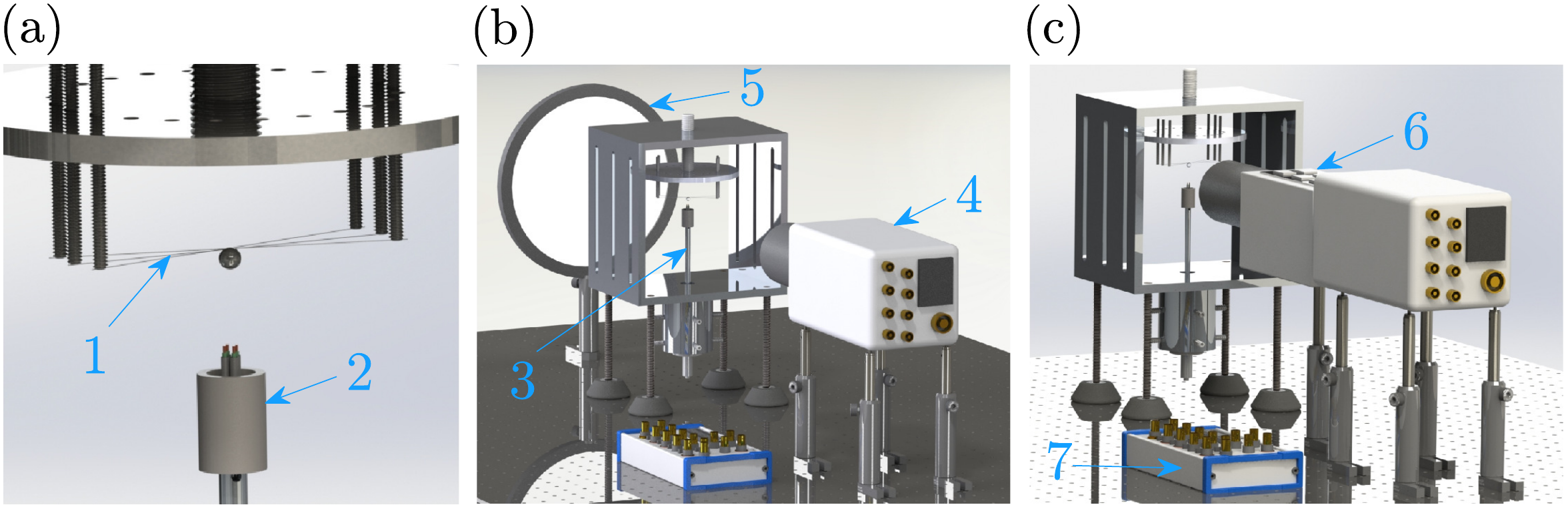}
	\caption{The experimental apparatus. (a) Close view of the cross-fiber configuration and the igniter, (b) the shadowgraphy setup, and (c) the $\mathrm{OH^*}$ chemiluminescence setup. The numbered items are: (1) cross-fiber, (2) igniter, (3) solenoid actuator, (4) high-speed camera, (5) back-light LED, (6) image intensifier, and (7) data acquisition unit. \label{fig:setdiag} }
\end{figure*}

The experimental setup is shown in Fig.~\ref{fig:setdiag}. Three $142~\mathrm{\mu m}$ diameter silicon carbide fibers (SCS-6, from Specialty Materials) were utilized as the suspension mechanism. As shown in Fig.~\ref{fig:setdiag}(a), the fibers (item 1) were placed inside a horizontal plane. Each fiber formed an angle of 15$^{\mathrm{o}}$ with the neighboring fiber. A gastight syringe (model 1710 from Harvard Apparatus) was used to suspend the fuel droplet at the fibers point of intersection using a syringe pump (11 Elite pump from Harvard Apparatus). The initial droplet diameter was $\mathrm{2.00}\pm\mathrm{0.02~mm}$. This yielded a droplet to fiber diameter ratio of about 14. Such relatively large ratio is expected to facilitate negligible heat transfer between the droplet and the suspension mechanism~\cite{avedisian2000soot}. A plasma-assisted igniter (item 2) was mounted on a linear actuator plunger (item~3 in Fig.~\ref{fig:setdiag}(b)) for ignition. For repeatability purposes, separate experiments were performed to ensure that the igniter arc is always formed at 1~mm below the droplet. The igniter operated for about 400~ms and was retracted using the actuator immediately after the operation period ended.

Separate high-speed shadowgraphy and $\mathrm{OH^*}$ chemiluminescence techniques were employed to study the droplet and flame dynamics. These diagnostics are shown in Figs.~\ref{fig:setdiag}(b) and (c), respectively. The acquisition frequency for both techniques was set to 4000~Hz. For both techniques, a high-speed camera (item 4 in the figure, Photron Fastcam Nova S12) was used. For the shadowgraphy experiments, the camera was equipped with a SIGMA Macro 105~mm ($f/\#=2.8$) lens. Similar to the studies of Bennewitz \textit{et al.}~\cite{bennewitz2020combustion,bennewitz2019systematic,bennewitz2018periodic}, a Semrock FF01--433/530 dual bandpass filter was installed on the camera lens to improve the visibility of the droplet in the presence of the diesel sooty flame. An LED light (item~5) was used for background illumination. The camera exposure time for the shadowgraphy technique was 100~$\mu$s. The imaging resolution was obtained using a USAF target plate following the procedure discussed in Papageorge \textit{et al.}~\cite{papageorge2014recent}, and the shadowgraphy resolution equals to $\mathrm{22~\mu m}/\mathrm{pixel}$. For the $\mathrm{OH^*}$ chemiluminescence measurements, an Invisible Vision UVi 1850B intensifier (item~6) equipped with a UV Nikon Rayfact 105~mm ($f/\# = 1.4$) lens was connected to the camera using a Nikkor 50~mm ($f/\# = 1.4$) lens. A $\mathrm{310}\pm\mathrm{10~nm}$ band-pass filter was installed on the UV lens to facilitate collecting flame chemiluminescence near the wavelength of the $\mathrm{OH^*}$ emission at 308~nm. The intensifier gate was set to 90~$\mu$s, and the pixel resolution of the chemiluminescence measurements was $\mathrm{23~\mu m/\mathrm{pixel}}$. All diagnostics were controlled and synchronized using an NI PCIe 6361 data acquisition board (item~7).
	
\subsection{Data reduction} \label{sec:reduction}
\begin{figure}[t!]
	\centering
	\includegraphics[width=0.5\textwidth]{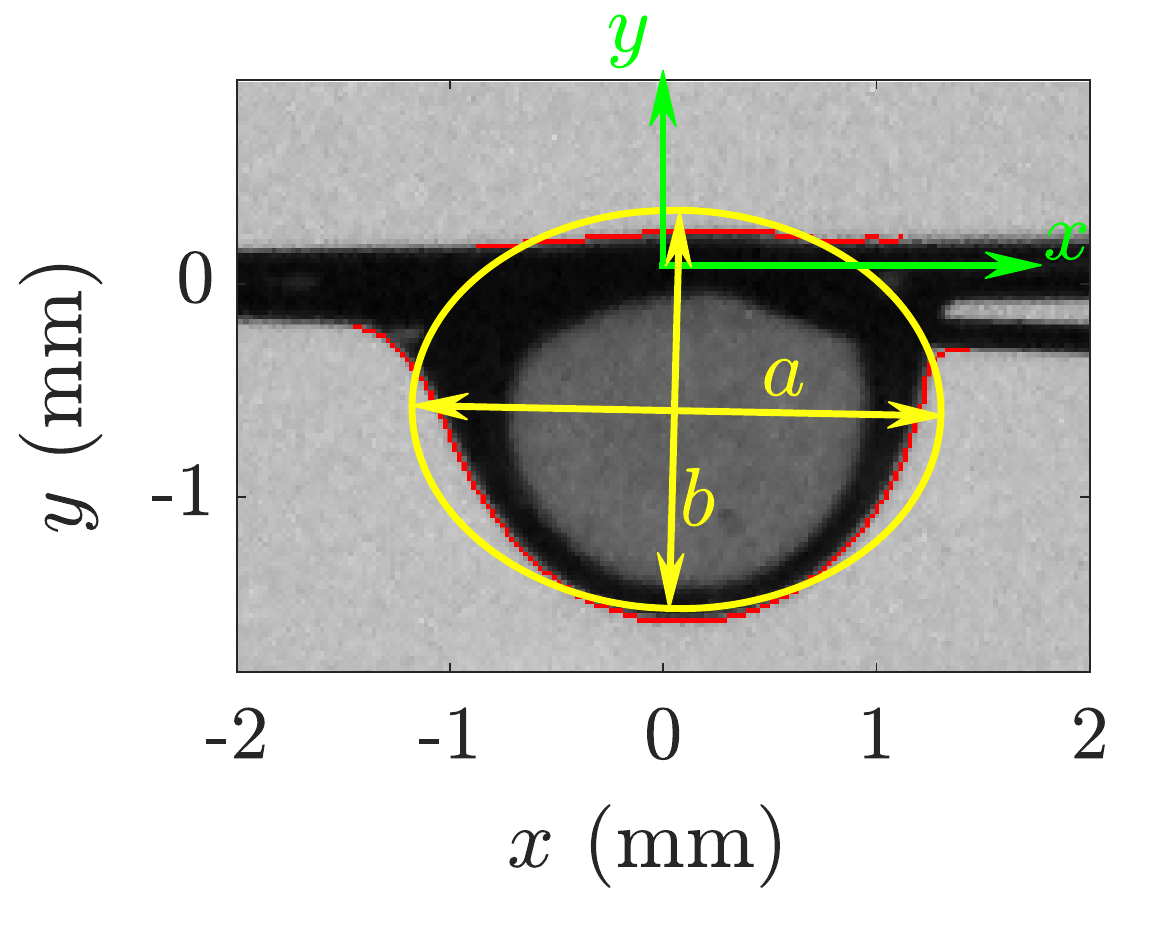}
	\caption{An ellipse with major and minor diameters of $a$ and $b$ fit to a shadowgraphy image of a droplet. The coordinate system is overlaid on the image.}
	\label{fig:datareduction}
\end{figure}

To estimate the droplet diameter, a procedure similar to that of~Hicks \textit{et al.}~\cite{hicks2010methanol}, Dietrich \textit{et al.}~\cite{dietrich2014droplet}, and Ghamari and Ratner~\cite{ghamari2017combustion} was followed. First, the droplet edges were obtained using the shadowgraphy technique. For a representative droplet, these edges are shown by the red color data points in Fig.~\mbox{\ref{fig:datareduction}}. Then, an ellipse with the major and minor diameters of $a$ and $b$ was fit to the droplet edges, see the yellow ellipse in Fig.~\ref{fig:datareduction}. Finally, the droplet equivalent diameter was calculated using $D=(ab^2)^{1/3}$. A Cartesian coordinate system was used to analyze the droplet dynamics. The $x$--axis is positioned on the wire centerline, and the $y$--axis is normal to the $x$--axis. The origin of the coordinate system was fixed and was selected such that $x = 0$ pertains to the center of the ellipse at $t = 0$.
	
It was not necessary to pre-process the shadowgraphy images for droplet dynamics analysis. However, the chemiluminescence images were pre-processed prior to their analyses. All acquired chemiluminescence images were subtracted by the background image (averaged for 200 chemiluminescence images before the droplet ignition) and were normalized for the spatially dependent sensitivity of the imaging equipment, similar to our past work~\cite{heydarlaki2021competing,mollahoseini121flame,mosadegh2021graphene} that utilized the same diagnostics. To minimize the electronics noise, a 7$\times$7~pixels$^2$ window median-based filter was applied to the chemiluminescence images.
	
	
	
\section{\label{sec:Results}Results and discussions}
The results are grouped into two subsections. First, the droplet diameter squared and the flame chemiluminescence oscillations as well as the atomization events characteristics are discussed. Then, in the second subsection, the coupling between the above parameters is elaborated. 

\subsection{Characteristics of the droplets, flames, and atomization events}
	
\begin{figure*}[t!]
\centering
\includegraphics[width=0.7\textwidth]{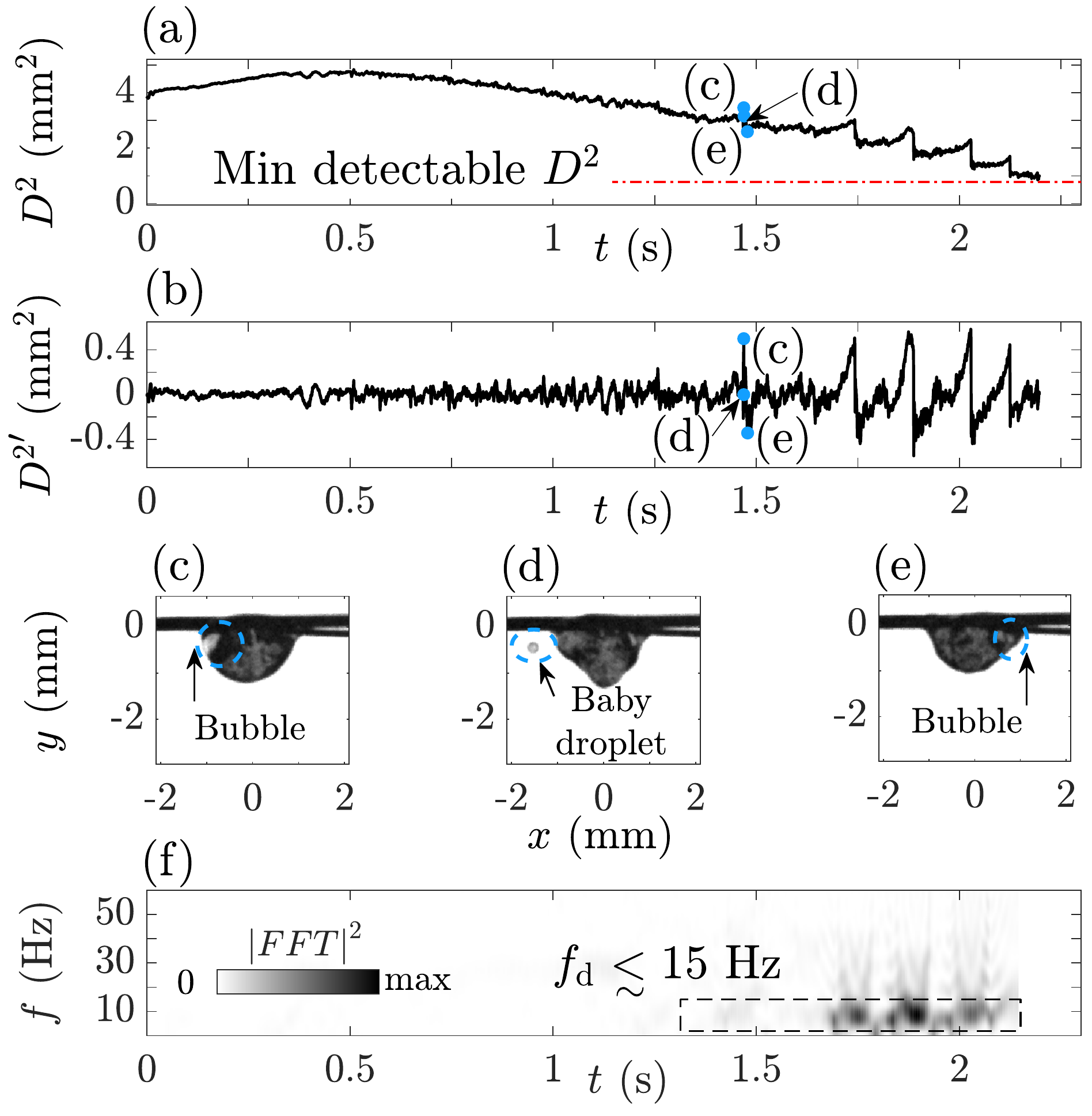}
\caption{(a) and (b) are $D^2$ and ${D^{2\prime}}$ versus $t$. (c--e) present the droplet images during an atomization event. (f) is the spectrogram of ${D^{2\prime}}$. 	\label{fig:droplet}}
\end{figure*}

Variations of the droplet diameter squared ($D^2$) versus time is presented in Fig.~\ref{fig:droplet}(a) for a representative test condition of D01. In addition to $D^2$, the present study is concerned with fluctuations of the droplet diameter squared ($D^{2\prime}$). Due to the inherently transient nature of the droplet combustion, a time-dependent mean diameter squared was obtained and subtracted from $D^2$ to estimate $D^{2\prime}$. That is $D^{2\prime} = D^2-\overline{D^2}$. The size of the time window for averaging the droplet diameter squared was selected to be large enough such that the droplet dynamics is resolved. The droplet diameter squared fluctuations corresponding to the results presented in Fig.~\ref{fig:droplet}(a) are shown in Fig.~\ref{fig:droplet}(b). As can be seen, $D^{2\prime}$ features relatively large amplitude oscillations. Three representative time instances corresponding to a large amplitude oscillation were selected and marked in Figs.~\ref{fig:droplet}(a and b), with the relevant droplet images shown in Figs.~\ref{fig:droplet}(c--e). As can be seen in Fig.~\ref{fig:droplet}(c), a bubble is present near the droplet surface immediately before its rupture. At this time instant, the bubble has reached its maximum size and the $D^2$ time series features a peak. After the main droplet surface is ruptured, a baby droplet is ejected, causing the main droplet surface corrugations as shown in Fig.~\ref{fig:droplet}(d). After about 10~ms that $D^{2\prime}$ featured a peak, this parameter is minimized and the next bubble appears at the droplet surface, see Fig.~\ref{fig:droplet}(e). The bubble detected in Fig.~\ref{fig:droplet}(e) leads to generation of the next baby droplet, and this process repeats.
	
The time-varying Fast Fourier Transform (FFT) of ${D^{2\prime}}$ (denoted by $|FFT|^2$) was calculated and is shown in Fig.~\ref{fig:droplet}(f). The size of the time window utilized for calculation of the spectrogram in Fig.~\ref{fig:droplet}(f) is selected using a trial and error process that leads to best presentation of the results. As can be seen in Fig.~\ref{fig:droplet}(f), the droplet diameter squared oscillations feature a dominant frequency of $f_\mathrm{d} \lesssim 15$~Hz. This frequency corresponds to the large amplitude fluctuations in Fig.~\ref{fig:droplet}(b).

	\begin{figure*}[h!]
	\centering
	\includegraphics[width=0.7\textwidth]{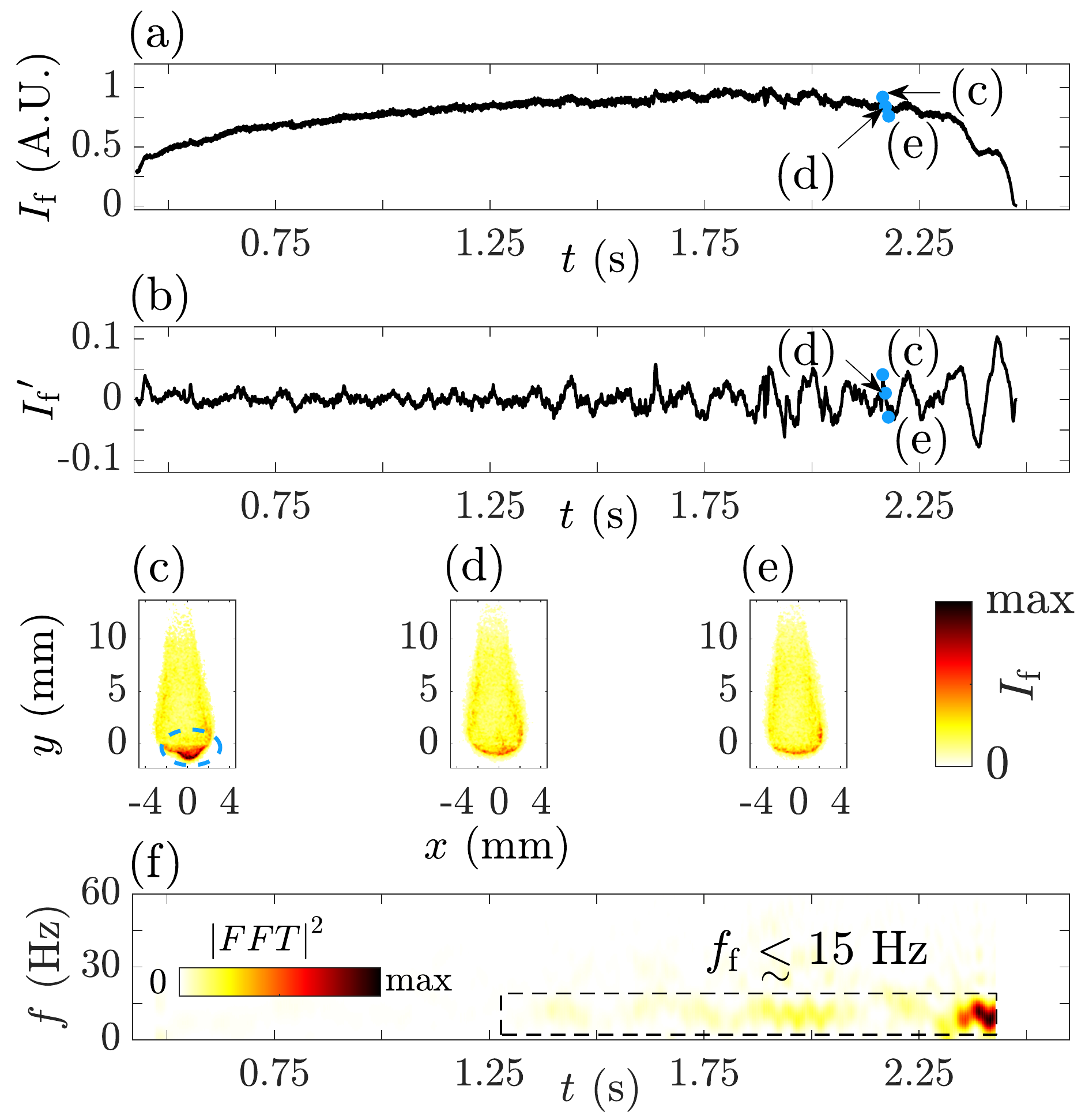}
	\caption{(a) and (b) are $I_\mathrm{f}$ and ${I_\mathrm{f}}^\prime$ versus $t$. (c--e) are the flame chemiluminescence during an atomization event. (f) is the spectrogram of ${I_\mathrm{f}}^\prime$. 
		\label{fig:flame}}
\end{figure*}

Variation of the flame chemiluminescence normalized by the corresponding maximum ($I_\mathrm{f}$) versus time for representative results pertaining to the test condition of D02 is presented in Fig.~\ref{fig:flame}(a). During the igniter operation period ($0~\mathrm{s} \leq t \lesssim 0.4~\mathrm{s}$), the chemiluminescence signal is contaminated by the igniter arc emission. Thus, the chemiluminescence data for $t \gtrsim 0.4~\mathrm{s}$ is presented and used for analysis. The normalized flame chemiluminescence intensity fluctuations signal (${I_\mathrm{f}}^\prime = I_\mathrm{f} - \overline{I}_\mathrm{f}$) is shown in Fig.~\ref{fig:flame}(b). A sliding window size identical to that used for obtaining the droplet diameter squared fluctuations was used for obtaining $I_\mathrm{f}^\prime$. Three time instants corresponding to a relatively large variation of $I_\mathrm{f}^\prime$ were selected and highlighted in Figs.~\ref{fig:flame}(a and b). The corresponding flame chemiluminescence intensity of these three time instants are presented in Figs.~\ref{fig:flame}(c--e). The relatively large value of ${I_\mathrm{f}}^\prime$ corresponding to Fig.~\ref{fig:flame}(c) is due to the ignition of ejected baby droplets and/or the entrainment of the fuel vapor into the flame area, as suggested by Pandey \textit{et al.}~\cite{pandey2019high}. Our flame chemiluminescence imaging confirmed the former. Indeed, analysis of the flame chemiluminescence videos show that once the baby droplets are ejected, they are ignited, and as a result, $I_\mathrm{f}$ increases to relatively large values. Once the baby droplets are burnt, the value of $I_\mathrm{f}$ decreases substantially, see Figs.~\ref{fig:flame}(d and e). Figure~\ref{fig:flame}(f) presents the spectrogram of $I_\mathrm{f}^\prime$ (denoted by $|FFT|^2$). As can be seen, similar to the results presented in Fig.~\ref{fig:droplet}(f), small frequency oscillations ($f_\mathrm{f} \lesssim 15$~Hz) are present towards the end of the droplet lifetime.
	
\begin{figure*}[h!]
\centering
\includegraphics[width=0.9\textwidth]{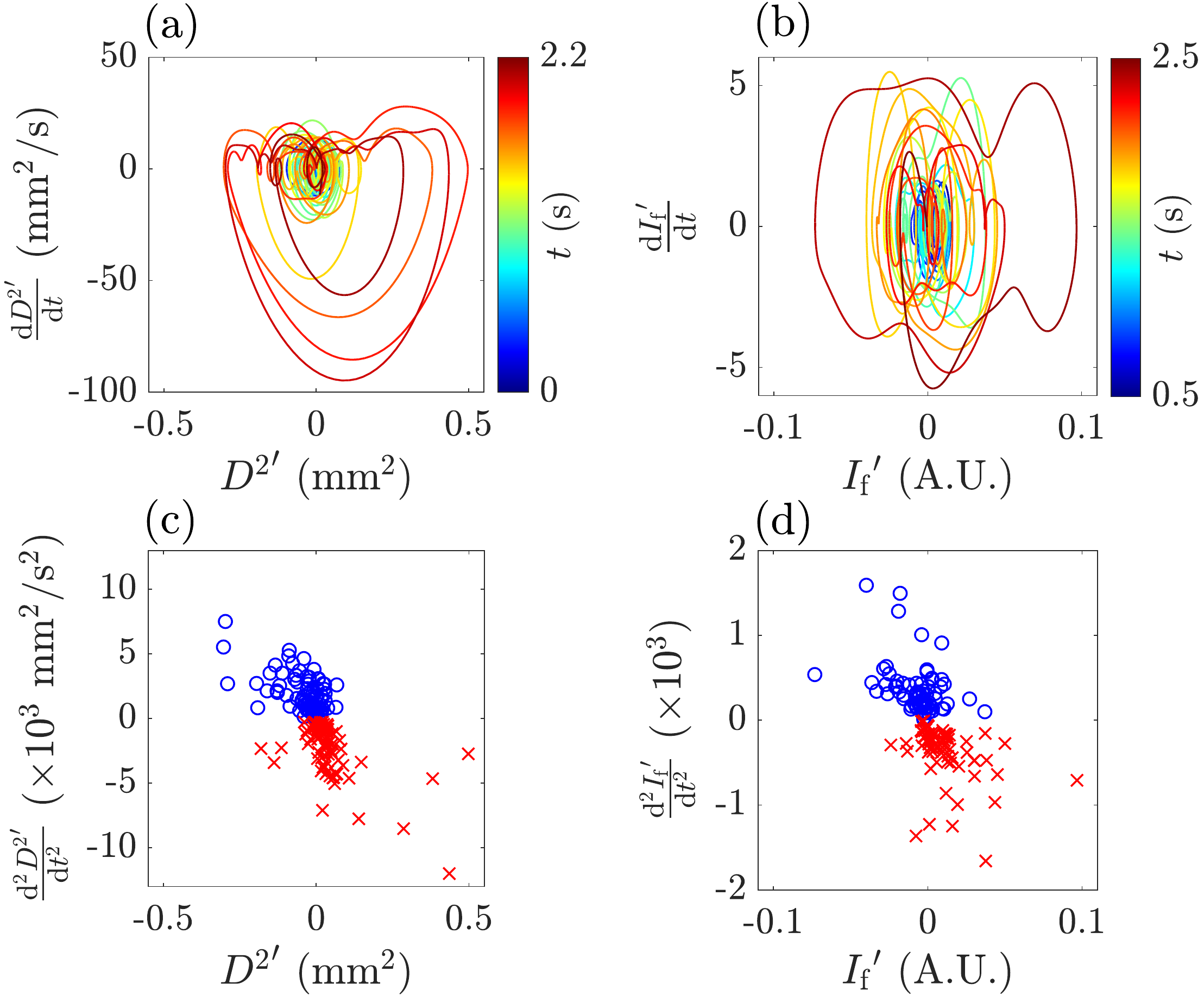}
\caption{(a) and (b) are the phase-space trajectories of the droplet and flame chemiluminescence oscillations. (c) and (d) are the corresponding Poincar\'{e} maps. The results pertain to D02 test condition.
}
\label{fig:Poincare}
\end{figure*}

The results presented in Figs.~\ref{fig:droplet}(b)~and~\ref{fig:flame}(b) suggest that the droplet diameter squared and flame chemiluminescence may feature intermittent oscillations. In order to asses this, the phase-space trajectories~\cite{kabiraj2015recurrence} and the Poincar\'{e} maps~\cite{heydarlaki2019influences} of $D^{2\prime}$ and $I'_\mathrm{f}$ were obtained and presented in Fig.~\mbox{\ref{fig:Poincare}}. Specifically, Figs.~\ref{fig:Poincare}(a)~and~(b) present variations of $D^{2\prime}$ and $I'_\mathrm{f}$ versus their corresponding first derivative with respect to time. The central differencing scheme~\cite{lomax2013fundamentals} was used to estimate the derivatives. The results in the figures are color coded based on time. As can be seen, the phase-space trajectories feature a large number of data points at their centers and some data points around their centers for both $D^{2\prime}$ and $I'_\mathrm{f}$. The second derivatives of $D^{2\prime}$ and $I'_\mathrm{f}$ were estimated and the data points at which the first derivatives of $D^{2\prime}$ and $I'_\mathrm{f}$ change sign were obtained. The collection of these data points is referred to as the Poincar\'{e} map~\cite{fountain2000chaotic}. These maps for the droplet diameter squared and flame chemiluminescence oscillations are presented in Figs.~\ref{fig:Poincare}(c)~and~(d), respectively. In the figures, the blue circular (red cross) data symbols correspond to those for which the first derivative of both $D^{2\prime}$ and $I'_\mathrm{f}$ change sign from negative (positive) to positive (negative). Agreeing with the phase-space trajectories, the Poincar\'{e} maps show that the data points are both clustered at the origin and scattered around the origin. The above analysis suggests that $D^{2\prime}$ and $I'_\mathrm{f}$ feature intermittent oscillations. This is similar to the characteristics of pressure and flame chemiluminescence oscillations presented in for example~Nair \textit{et al.}~\cite{sujith_2014}, Kheirkhah \textit{et al.}~\cite{kheirkhah2017dynamics}, Guan \textit{et al.}~\cite{guan2020intermittency}, and Ma \textit{et al.}~\cite{ma2022interpretation}. Although the results corresponding to test the condition of D02 are presented here, indeed, similar intermittent behavior was also observed for the rest of the tested conditions. The intermittent oscillations of $D^{2\prime}$ and $I'_\mathrm{f}$ are observed, characterized, and reported for the first time in the present study.
	
\begin{figure*}[h!]
\centering
\includegraphics[width=1\textwidth]{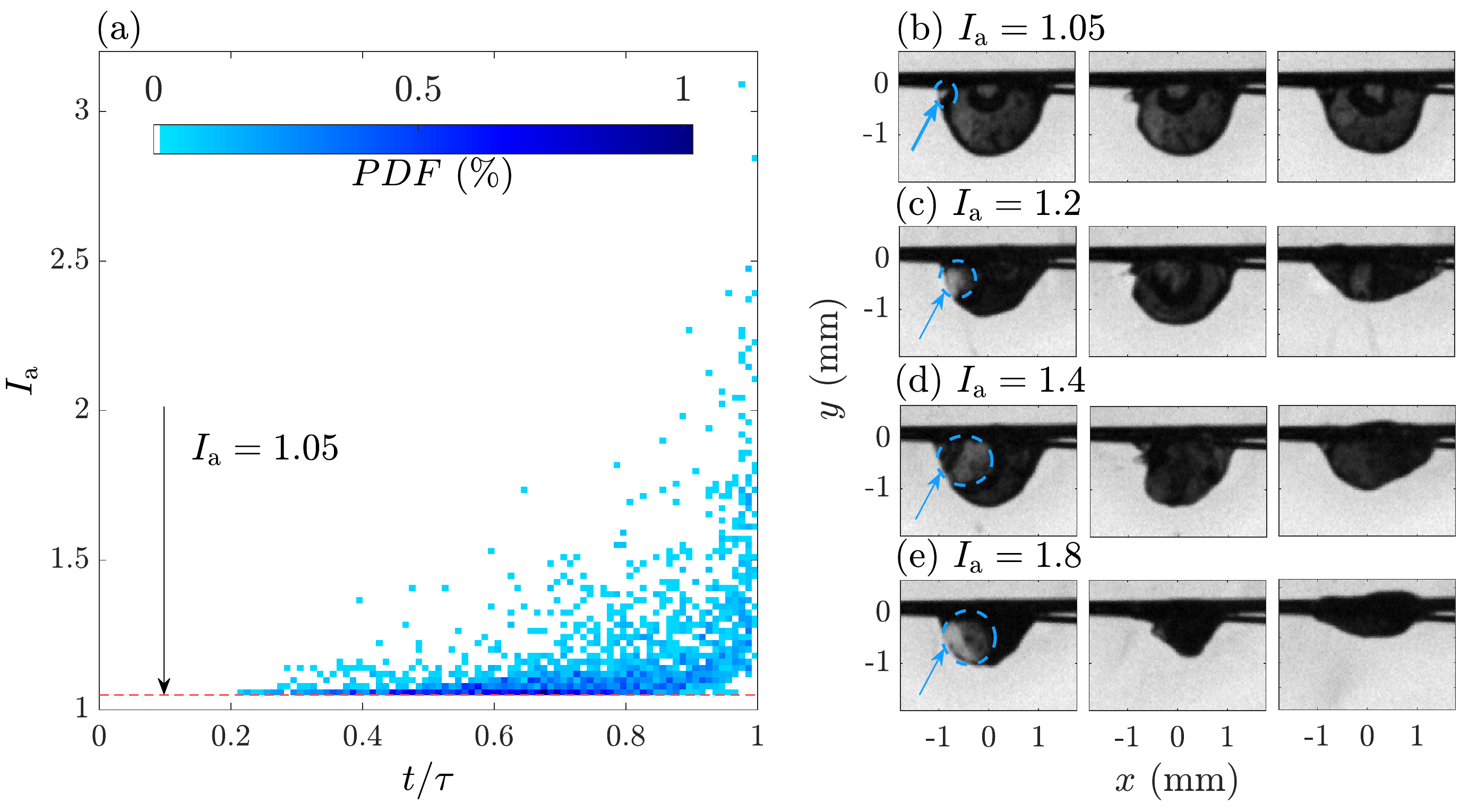}
\vspace{1 pt}
\caption{(a) Time dependent PDF of the atomization intensity, $I_\mathrm{a}$. (b--e) are the representative shadowgraphy images during atomization events with $I_\mathrm{a} = 1.05$, 1.2, 1.4, and 1.8, respectively.
\label{fig:IaPDF}} 
\end{figure*}

It is hypothesized that the similarity between the characteristics of the droplet diameter squared and the flame chemiluminescence oscillations (presented in Figs.~\ref{fig:droplet}--\ref{fig:Poincare}) is not coincidental, and it is due to the occurrence of the atomization events. In order to characterize these events as well as to assess the above hypothesis, the $D^2$ time series, similar to that shown in Fig.~\ref{fig:droplet}(a), are used. Analysis of the $D^2$ data as well as the shadowgraphy images shows that the beginning and the end of an atomization event correspond to a relatively large variation in the $D^2$ values, as evident in Figs.~\ref{fig:droplet}(c--e). Here, the atomization intensity ($I_\mathrm{a}$) is defined as $I_\mathrm{a}=D_\mathrm{i}^2/D_\mathrm{e}^2$, with $D_\mathrm{i}^2$ and $D_\mathrm{e}^2$ being the droplet diameter squared immediately at the beginning and at the end of an atomization event, respectively. The droplet diameter squared primarily changes due to the atomization and evaporation. Analysis of $D^2$ data suggests that the smallest value of $D^2$ that can be detected is 1~$\mathrm{mm^2}$ (see the red dashed line in Fig.~\ref{fig:droplet}(a)) due to the presence of the droplet suspension mechanism; the maximum duration of an atomization event is $\sim$30~ms; and that $|\mathrm{d}D^2/\mathrm{d}t|$ features a maximum of 1.5~$\mathrm{mm^2/s}$ in the absence of atomization. This means that the percentile change in the droplet diameter squared due to evaporation features a maximum of about $(1.5~\mathrm{mm^2/s}~\times~0.03~\mathrm{s})/1~\mathrm{mm^2}~\approx 5\%$. As a result, events with $I_\mathrm{a} \gtrsim 1.05$ during time periods smaller than 30~ms are expected to correspond to the atomization. Indeed, using the shadowgraphy images, this was confirmed that $D_\mathrm{i}^2$ is larger than $D_\mathrm{e}^2$ by at least 5\% during an atomization event. The values of $I_\mathrm{a}$ were obtained for the 12 repeats of all tested conditions, the Probability Density Function ($PDF$) of this parameter normalized by the corresponding maximum was obtained, and the results are presented in Fig.~\ref{fig:IaPDF}(a). In the figure, the PDF was calculated for several temporal bins, aiming to highlight how the intensity of atomization change during time. As can be seen, $I_\mathrm{a}$ can become as large as about 3; and, the atomization events mostly occur towards the end of the droplet lifetime. This agrees with the results presented in Figs.~\ref{fig:droplet}(f)~and~\ref{fig:flame}(f) and suggests there may exist a causal relation between the atomization events and the droplet diameter squared as well as the flame chemiluminescence oscillations. For demonstration purposes, representative shadowgraphy images corresponding to representative events with $I_\mathrm{a} = 1.05$, $1.2$, $1.4$, and $1.8$ are shown in the series of images in Figs.~\ref{fig:IaPDF}(b--e), respectively. As can be seen, all images show the presence of bubbles prior to rupturing of the droplet surface (see the blue circles and arrows). This confirms that, indeed, $I_\mathrm{a} \geq 1.05$ allows for detection of the atomization events. While this criterion can change for other doped fuels and shadowgraphy settings, it allows for detection of the atomization events in the present study.
	
\begin{figure*}[h!]
\centering
\includegraphics[width=0.8\textwidth]{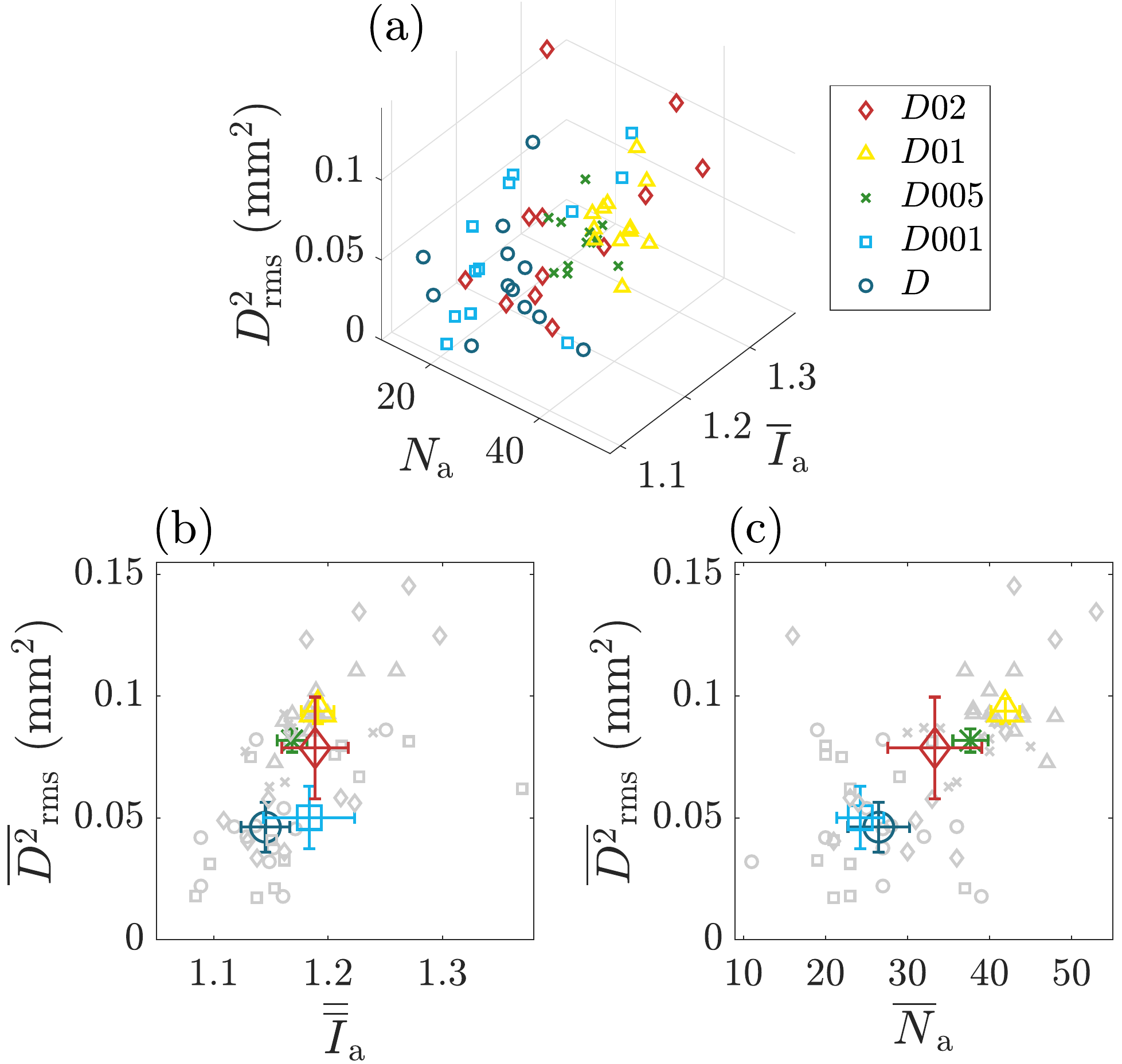}
\caption{(a) The RMS of the $D^{2\prime}$ fluctuations versus the number of atomization events and the mean atomization intensity for all tested conditions and their repeats. (b) and (c) are the 2D presentation of the data in (a), with the colored and enlarged data points presenting the averaged values. \label{fig:rms3d}}
\end{figure*}

The relation between the atomization events and the droplet diameter squared RMS was investigated statistically. Figure~\ref{fig:rms3d}(a) presents variation of $D^{2\prime}_\mathrm{rms}$ versus both the mean atomization intensity ($\overline{I}_\mathrm{a}$) as well as the number of the atomization events ($N_\mathrm{a}$) for all tested conditions and their repeats. For clarity purposes, variation of $D^{2\prime}_\mathrm{rms}$ averaged for all repeats of each tested condition (denoted by $\overline{D^{2\prime}}_\mathrm{rms}$) versus the mean atomization intensity also averaged for the repeats of each tested condition ($\overline{\overline{I}}_\mathrm{a}$) is presented in Fig.~\ref{fig:rms3d}(b). Variation of $\overline{D^{2\prime}}_\mathrm{rms}$ versus the number of the atomization events averaged for the repeats of each tested condition ($\overline{N}_\mathrm{a}$) is presented in Fig.~\ref{fig:rms3d}(c). The lengths of the error bars in Figs.~\ref{fig:rms3d}(b and c) present the standard deviations of the relevant parameters. The data shown by the light gray color in Figs.~\ref{fig:rms3d}(b and c) represent the variability due to the repeats of the tested conditions. As can be seen, generally, increasing the number and intensity of the atomization events increase the RMS of the droplet diameter squared oscillations. The results in the figure show that increasing the doping concentration from 0.001 to 0.02\% increases $\overline{D^{2\prime}}_\mathrm{rms}$ by about 2 folds. Although the number of atomization events and their intensity are influenced by the doping concentration, it is unclear if the droplet and flame dynamics are influenced by the atomization events at all. Such coupling is discussed in the next subsection.
	
\subsection{The coupling between the droplets, flames, and atomization events}

\begin{figure*}[h!]
\centering
\includegraphics[width=1\textwidth]{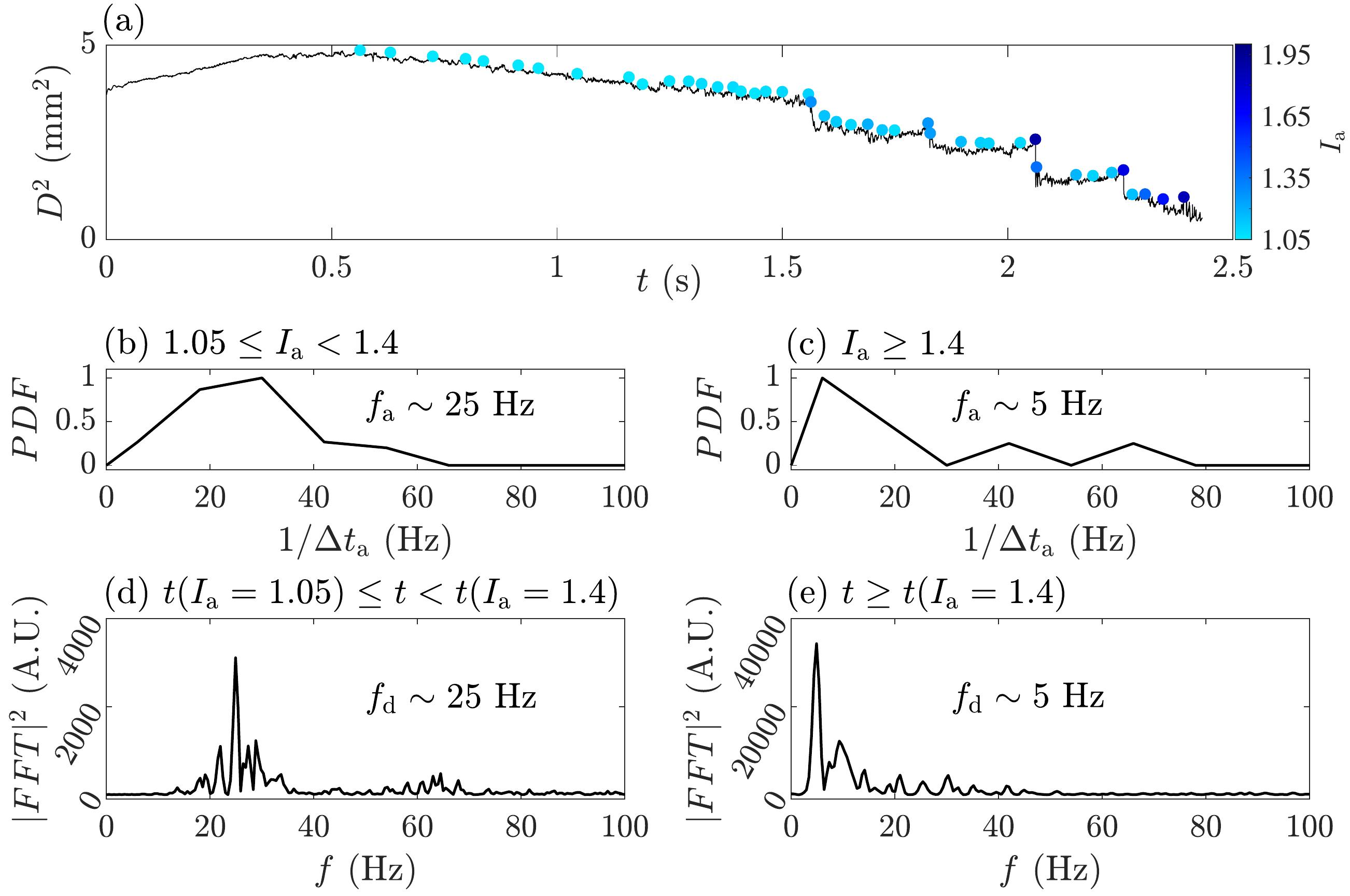}
\caption{(a) Variation of $D^2$ versus time. The results correspond to D01 test condition. (b) and (c) are the probability density functions of the inverse of the time between to consecutive atomization events with $1.05 \leq I_\mathrm{a} < 1.4$ and $I_\mathrm{a} \geq 1.4$, respectively. (d) and (e) are the PSDs of the droplet diameter squared oscillations corresponding to $t(I_\mathrm{a} = 1.05) \leq t < t(I_\mathrm{a} = 1.4)$ and $t \geq t(I_\mathrm{a} = 1.4)$, respectively.}
\label{fig:atomcurve}
\end{figure*}

For a representative test condition of D01, the variation of $D^2$ versus time is presented in Fig.~\ref{fig:atomcurve}(a). Overlaid on the figure are the solid circular data points that correspond to the atomization events, which are highlighted for $I_\mathrm{a} \geq 1.05$. These data points are also color-coded based on the value of $I_\mathrm{a}$, i.e. dark blue corresponds to intense atomization events. As can be seen, generally, large intensity atomization events occur towards the end of the droplet lifetime, which confirms the results presented in Fig.~\ref{fig:IaPDF}(a).
	
It is of interest to investigate if there exists a relation between the intensity of the atomization events and the time separation between two consecutive atomization event, referred to as $\Delta t_\mathrm{a}$. In order to study this, the atomization events were categorized using a threshold value of $I_\mathrm{a} = 1.4$, and the PDF of $1/\Delta t_\mathrm{a}$ was calculated. For the representative test condition of D01, the PDFs of $1/\Delta t_\mathrm{a}$ for atomization events with $1.05 \leq I_\mathrm{a} < 1.4$ and $I_\mathrm{a} \geq 1.4$ are presented in Figs.~\ref{fig:atomcurve}(b) and (c), respectively. The results in Fig.~\ref{fig:atomcurve}(c) suggest that, for $I_\mathrm{a} \geq 1.4$, the atomization events become relatively scarce such that a best presentation of the PDF is not possible. The results show that the most probable values of $1/\Delta t_\mathrm{a}$ occur at about 25 and 5~Hz for $1.05 \leq I_\mathrm{a} < 1.4$ and $I_\mathrm{a} \geq 1.4$, respectively. These dominant frequencies were calculated using the area-weighted average of the probability density function for $PDF>0.5\times\max\{PDF\}$. The results presented in Figs.~\ref{fig:atomcurve}(a--c) suggest that, for this tested condition, atomization event with large intensities occur less frequently and towards the end of the droplet lifetime.

The results in Fig.~\ref{fig:atomcurve}(a) as well as the atomization intensity were used to identify the moments corresponding to the first time at which $I_\mathrm{a} = 1.05$ and 1.4. These are referred to as $t(I_\mathrm{a} = 1.05)$ and $t(I_\mathrm{a} = 1.4)$. Results in Figs.~\ref{fig:atomcurve}(d) and (e) present the power spectrum densities (PSD) of the droplet diameter squared oscillations for $t(I_\mathrm{a} = 1.05) \leq t < t(I_\mathrm{a} = 1.4)$ and $t \geq t(I_\mathrm{a} = 1.4)$, respectively. As can be seen, during the the time period that the atomization events are not intense, the PSD of the droplet diameter squared oscillations feature a dominant frequency near 25~Hz, which is similar to that reported in Fig.~\ref{fig:atomcurve}(b). However, for the rest of the droplet lifetime (corresponding to intense atomization events) the droplet diameter squared oscillations feature a dominant frequency near 5~Hz, which is similar to that reported in Fig.~\ref{fig:atomcurve}(c). The above argument suggests a causality relation between the atomization events and the droplet diameter squared oscillations.

It is noted that the amplitude of the PSD presented in Fig.~\ref{fig:atomcurve}(e) and at $f \approx 5$~Hz is about one order of magnitude larger than that for the PSD presented in Fig.~\ref{fig:atomcurve}(d) at $f \approx 25$~Hz. As discussed in Appendix A, the oscillations at 25~Hz are also present prior to the occurrence the first atomization event, are due to the droplet vibrations along the vertical axis, and are originally induced as a result of the retracting motion of the igniter.

\begin{figure*}[h!]
\centering
\includegraphics[width=0.6\textwidth]{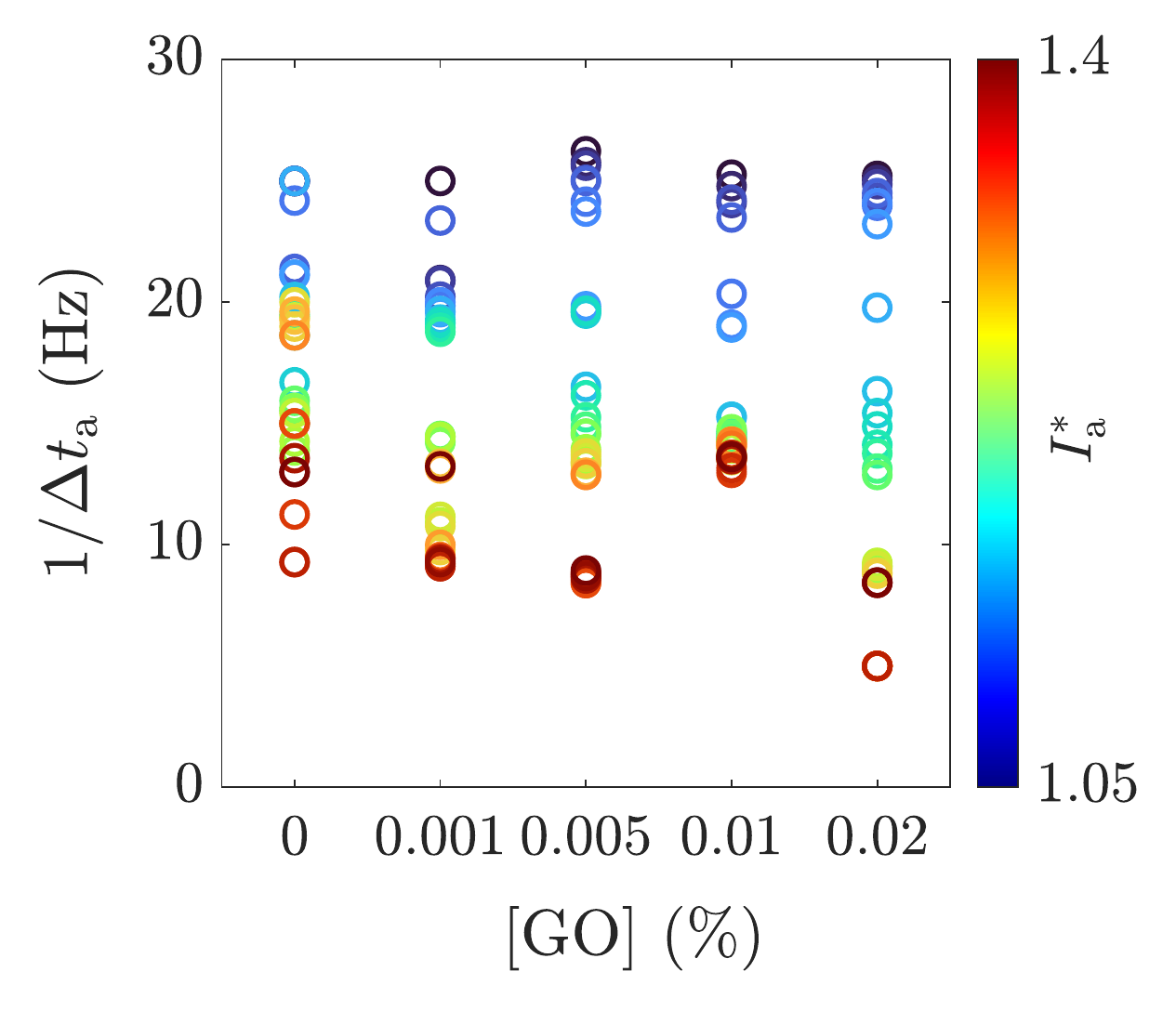}
\caption{The most probable value of $1/\Delta t_\mathrm{a}$ for atomization events with $I_\mathrm{a} \geq I^*_\mathrm{a}$. \label{fig:fIa}}
\end{figure*}

It is of interest to confirm if the observations presented in Fig.~\ref{fig:atomcurve} hold for all tested conditions. Following the above discussions, the PDF of $1/\Delta t_\mathrm{a}$ was estimated for $I_\mathrm{a} \geq I^*_\mathrm{a}$, with $I^*_\mathrm{a}$ being a threshold atomization intensity value. For all tested conditions, the most probable value of $1/\Delta t_\mathrm{a}$ for several values of $ 1.05 \leq I^*_\mathrm{a} \leq 1.4$ were obtained, averaged for 12 repeats of each test condition, and presented in Fig.~\ref{fig:fIa}. The results in the figure confirm that the frequency of occurrence of the intense atomization events decrease as $I^*_\mathrm{a}$ increases, and this is consistent for all tested conditions.
	
\begin{figure*}[h!]
\centering
\includegraphics[width=\textwidth]{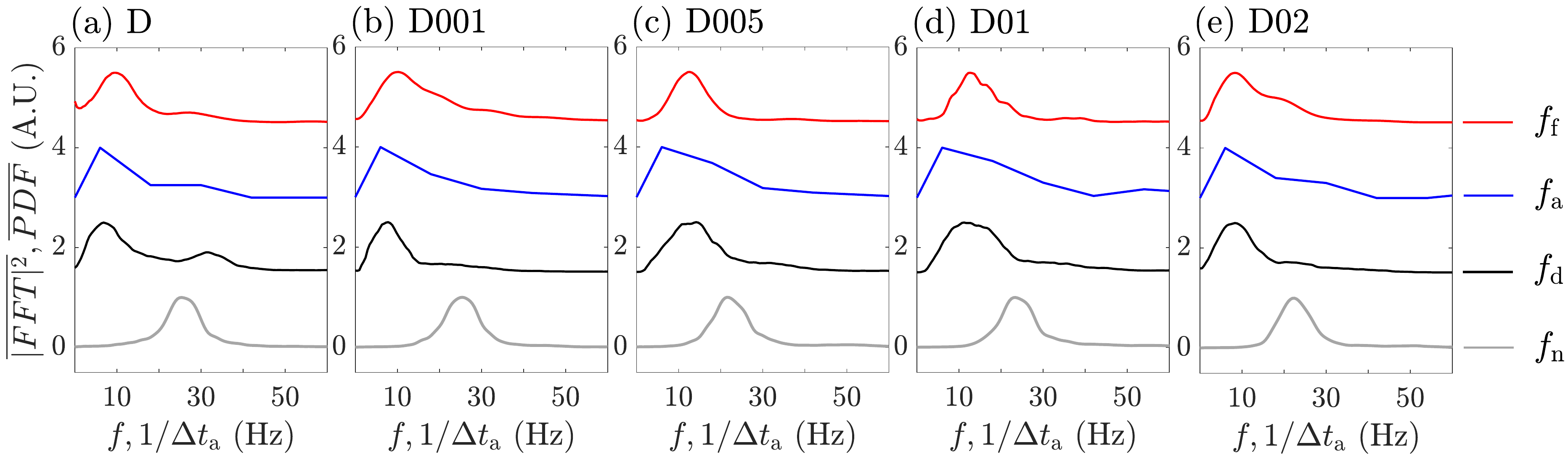}
\caption{Spectra of the droplet diameter squared (black) and flame chemiluminescence (red) oscillations as well as the PDF of the inverse of the time difference between two consecutive atomization events (blue). The gray curves are the spectra of the droplet vibrations along the vertical axis. \label{fig:FFT}}
\end{figure*}

\begin{figure*}[h!]
\centering
\includegraphics[width=0.7\textwidth]{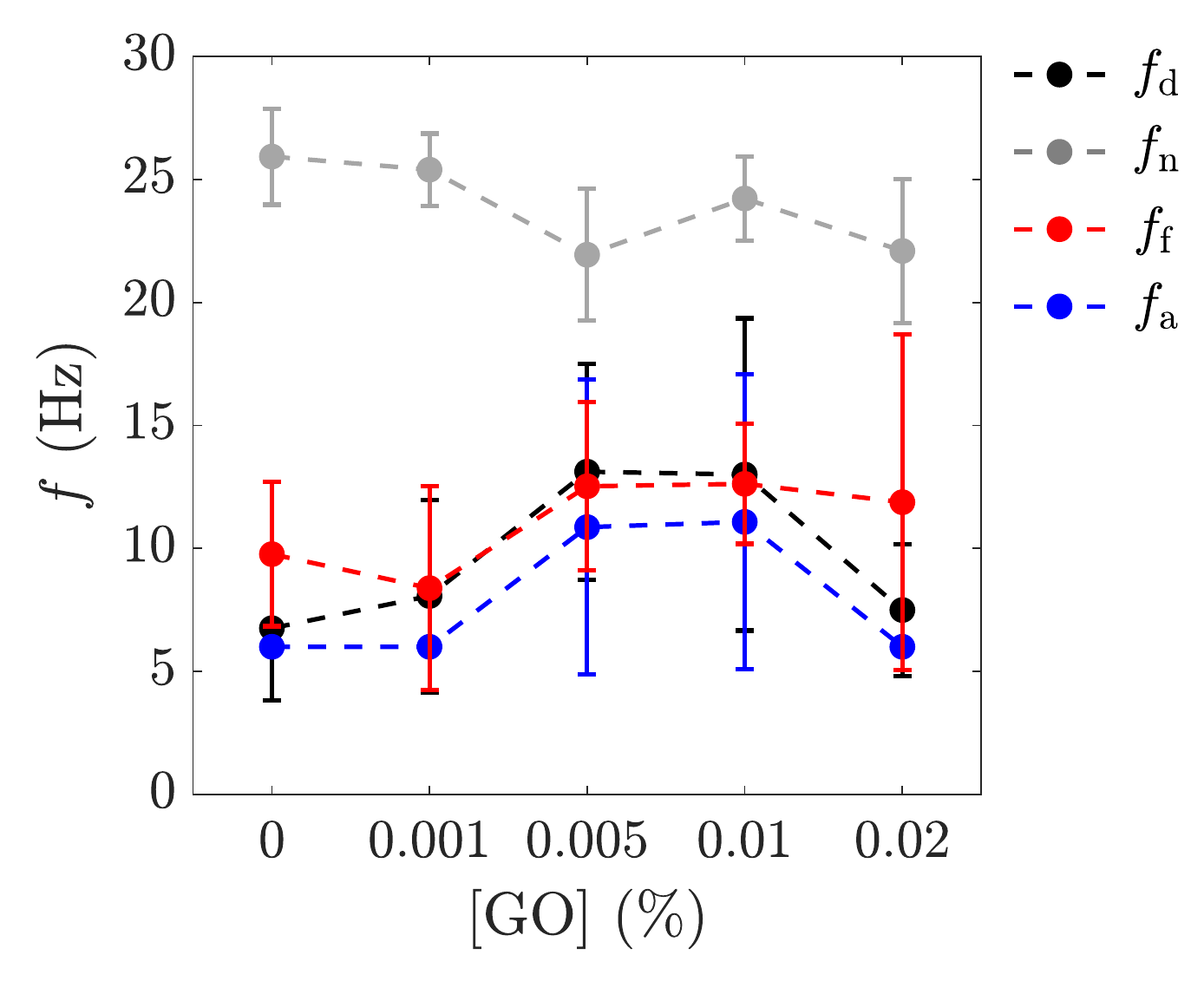}
\caption{Variations of the dominant frequencies of the droplet oscillations along the vertical axis ($f_\mathrm{n}$), flame chemiluminescence oscillations ($f_\mathrm{f}$), and the droplet diameter squared oscillations after the occurrence of the first intense atomization event ($f_\mathrm{d}$). $f_\mathrm{a}$ is the most probable frequency of intense atomization events. \label{fig:frequencies}}
\end{figure*}

Although the above discussions elaborate the connection between the atomization events and the droplet diameter squared fluctuations, it is unclear how these relate to the flame chemiluminescence oscillations, specifically for intense atomization events. In order to investigate this, the time instant at which the first atomization event with $I_\mathrm{a} \geq 1.4$ (i.e. the intense atomization) occurs was obtained. The power spectrum density of the droplet diameter squared fluctuations for the time period immediately after the occurrence of the first intense atomization event is presented in Fig.~\ref{fig:FFT}, see the black curves. The results in Figs.~\ref{fig:FFT}(a--e) pertain to the test conditions of D, D001, D005, D01, and D02, respectively. The dominant frequencies for the droplet diameter squared oscillations were extracted and presented in Fig.~\ref{fig:frequencies}, see the black data symbol in the figure. The error bars in the figure are the Full Width at Half Maximum of the PSDs. The PSDs of the flame chemiluminescence oscillations were also calculated for all tested conditions and are presented by the solid red curve in Fig.~\ref{fig:FFT}. The dominant frequency of the flame chemiluminescence oscillations was extracted from the results in Fig.~\ref{fig:FFT} and are presented in Fig.~\ref{fig:frequencies} using the red data points. For comparison purposes, the PDF of the inverse of the time difference between the occurrences of two consecutive atomization events (for $I_\mathrm{a} \geq 1.4$) as well as the PSD of the oscillation of the droplet center along the vertical axis (which is also presented in Appendix A) are shown by the blue and gray curves in Fig.~\ref{fig:FFT}, respectively. The results in Fig.~\ref{fig:FFT} are stepped for clarity purposes. The dominant frequency of the vibrations spectrum ($f_\mathrm{n}$) and the most probable value of intense atomization events frequency ($f_\mathrm{a}$) are shown by the gray and blue data points in Fig.~\ref{fig:frequencies}. As can be seen, for all tested conditions, the dominant frequencies of the droplet diameter squared oscillations spectrum, the most probable frequency of intense atomizations, and the dominant frequency of the flame chemiluminescence oscillations spectrum match. The results show that $f_\mathrm{n}$ is appreciably larger than $f_\mathrm{d}$, $f_\mathrm{a}$, and $f_\mathrm{f}$. It is also observed that, although addition of GO can change the atomization intensity (as shown earlier in Fig.~\ref{fig:rms3d}), $f_\mathrm{n}$, $f_\mathrm{d}$, $f_\mathrm{a}$, and $f_\mathrm{f}$ are not influenced by GO addition.
	
\section{Concluding remarks} \addvspace{10pt}
\label{Sec:Conclusions}
The characteristics as well as the coupling between the droplet diameter squared and flame chemiluminescence oscillations as well as atomization events were studied experimentally. Both neat and graphene oxide doped diesel droplets, with doping concentrations by weight of 0.001, 0.005, 0.01, and 0.02\%, were utilized. All droplets were suspended at the intersection point of three fibers, with the fiber diameter of $142~\mu$m. Separate shadowgraphy and $\mathrm{OH^*}$ chemiluminescence measurements were performed to study the droplet and the flame dynamics, respectively. The data acquisition for both techniques was set to 4000~Hz.
	
The results showed that the atomization was present for all tested conditions, and it induced both the droplet diameter squared and flame chemiluminescence oscillations. These oscillations were shown to feature intermittent behaviour, demonstrated by the phase-space trajectories and Poincar\'{e} maps. The intensity of the atomization events was quantified using the ratio of the droplet diameter squared at the beginning of each event to that at the end. The time varying PDF of the atomization intensity suggested that the possibility of large intensity atomization events increased towards the end of the droplet lifetime. It was shown that increasing the mean number and intensity of the atomization events increased the RMS of the droplet diameter squared oscillations, with the doping concentrations of 0.005--0.02\% featuring about 100\% larger RMS values compared to those with smaller doping concentrations.
	
The atomization intensity was used to condition the droplet diameter squared oscillations. Prior to the occurrence of the intense atomization events, both the probability density function of the inverse of the time separation between two consecutive atomization events and the power spectrum density of the droplet diameter squared oscillations featured large probabilities and powers at about 25~Hz. After the occurrence of the intense atomization events, this frequency decreased to about 5~Hz for both. The power spectrum density of the droplet diameter squared oscillations at 5~Hz was about one order of magnitude large than that at 25~Hz. It was shown that, while the atomization caused the large amplitude oscillations at 5~Hz, the retracting motion of the igniter induced the oscillations at 25~Hz. 

For all tested conditions, the dominant frequencies of the flame chemiluminescence, droplet diameter squared, the vertical position of the droplet center, and the atomization events were obtained. It was observed that, depending on the doping concentration, the droplet vertical oscillations frequency can vary from about 21 to 25~Hz, but the dominant frequencies of the rest vary from about 5 to 13~Hz. In essence, for all doping concentrations, the droplet diameter squared and flame chemiluminescence oscillations frequencies lock-on to that of the intense atomization events, which are relatively small (less than 15~Hz). Such small frequency coupling between the droplets, their flames, and atomizations is of relevance to spray combustion research, which is the subject of future investigations.
	
\textbf{Acknowledgments}

The authors are grateful for financial support from the Natural Sciences and Engineering Research Council (NSERC) Canada and Zentek through the Collaborative Research and Development grant (CRDPJ 536828--18). The authors are thankful to Dr. Colin van der Kuur of Zentek for discussions.

\appendix
	
\section{Droplet oscillations along the vertical axis}

\begin{figure}[h!]
	\centering
	\includegraphics[width=1\textwidth]{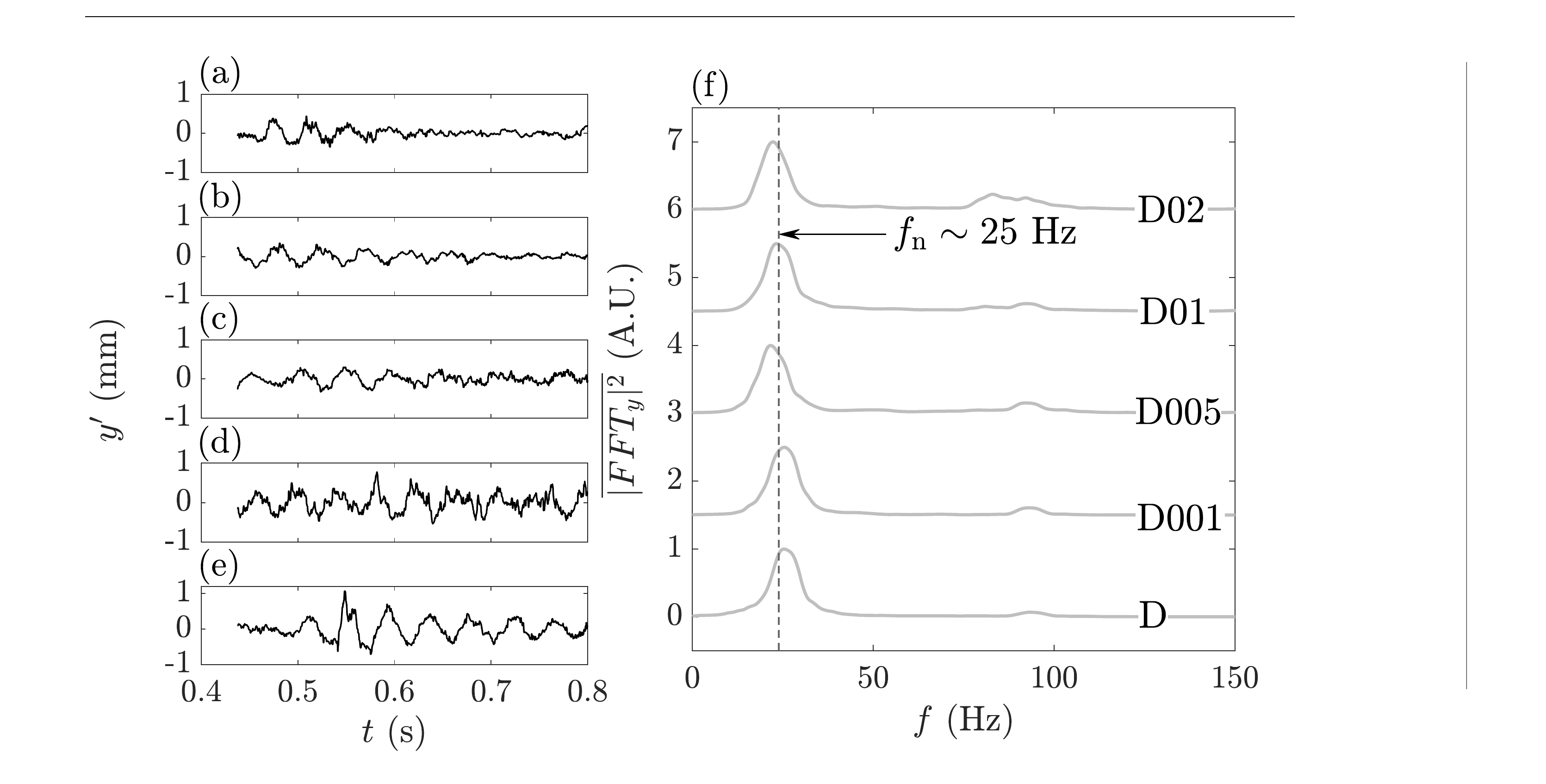}
	\vspace{1 pt}
	\caption{(a--e) are representative oscillations of the droplet center along the vertical axis versus time for test conditions of D, D001, D005, D01, and D02, respectively. (f) presents the averaged power spectrum densities of the these oscillations. \label{fig:naturalfrequency}}
\end{figure}

The dynamics of the droplet oscillations along the vertical axis is discussed in this appendix. It was confirmed that the wires used to suspend the droplets did not move/vibrate for all tested conditions and their repeats. However, compared to the wires, droplets could oscillate primarily along the vertical axis. It was observed that the retracting motion of the igniter, which occurs at $t = 0.4$~s, induced the droplet oscillations. These oscillations are present prior to the first atomization event and are quantified using the vertical position of the droplet center. It is claimed that the droplet oscillations with respect to the wires are not the root cause of a potential coupling between the droplet diameter squared and the flame chemiluminescence oscillations. As postulated in Section~\ref{sec:Results}, indeed, the atomization is the root cause of the above coupling. Representative variations of the droplet center vertical position oscillations versus time for test conditions of D, D001, D005, D01, and D02 are presented in Figs.~\ref{fig:naturalfrequency}(a--e), respectively. The power spectrum densities of the oscillations of the droplet along the vertical axis were calculated for all 12 repeats of each tested conditions. The power spectrum densities were then averaged and presented in Fig.~\ref{fig:naturalfrequency}(f). The PSDs in the figure were stepped for clarity purposes. As can be seen, for all tested conditions, the droplets move at a frequency of about 25~Hz. As discussed in Section~\ref{sec:Results}, this frequency is appreciably larger than that associated with the droplet diameter squared oscillations, flame chemiluminescence oscillations, and the atomization events. The above argument suggests that the root cause of the coupling is not due to the droplet oscillations. However, the retracting motion of the igniter is the root cause of the 25~Hz oscillations prior to the occurrence of the first intense atomization event.

\bibliography{aipsamp}
	
\end{document}